\documentstyle[12pt]{article}
\newcommand{\MeV}{{\rm MeV}}                    
\newcommand{\fm}{{\rm fm}}                      
\newcommand{\BUU}{{\sl BUU}}                    
\newcommand{\BL}{{\sl BL}}                      
\newcommand{\rme}{{\rm e}}                      
\newcommand{\del}{\partial}
\newcommand{\vareps}{\varepsilon}
\newcommand{\eps}{\epsilon}

\newcommand{\bold}[1]{{\mbox{\boldmath $#1$}}}    
\newcommand{\r}{\bold{r}}			
\newcommand{\p}{\bold{p}}                       
\newcommand{\k}{{\bf k}}			
\newcommand{\bfv}{\bold{v}}                    
\newcommand{\ie}{{\em i.e.}}                    
\newcommand{\etal}{{\em et al.}}                
\newcommand{\beq}{\begin{equation}}
\newcommand{\eeq}{\end{equation}}
\newcommand{\beqar}{\begin{eqnarray}}
\newcommand{\eeqar}{\end{eqnarray}}
\newcommand{\bd}{\begin{itemize}} 
\newcommand{\ed}{\end{itemize}} 
\newcommand{\bc}{\begin{center}}
\newcommand{\ec}{\end{center}}

\newcommand{\be}{\begin{equation}}
\newcommand{\ee}{\end{equation}}
\newcommand{\ba}{\begin{array}}
\newcommand{\ea}{\end{array}}
\newcommand{\bfig}{\begin{figure}}
\newcommand{\efig}{\end{figure}}

\begin{document}
\begin{titlepage}
\noindent{\sl Physical Review C}
\hfill LBL-35987\\[8ex]

\begin{center}
{\large {\bf
Analysis of Boltzmann-Langevin dynamics\\ in nuclear matter$^*$}}\\[8ex]
{\sl S. Ayik$^1$, Ph. Chomaz$^2$, M. Colonna$^{2,3}$,
and J. Randrup$^4$}\\[5ex]

$^{1)}$Department of Physics, Tennessee Technological University\\
Cookeville, Tennessee 38505\\[2ex]

$^{2)}$ GANIL, B.P. 5027, F-14021 Caen Cedex, France\\[2ex]

$^{3)}$ LNS, Viale Andrea Doria, Catania, Italy\\[2ex]

$^{4)}$ Nuclear Science Division, Lawrence Berkeley Laboratory\\
University of California, Berkeley, California 94720, USA\\[4ex]

\today\\[4ex]
{\sl Abstract}
\end{center}
The Boltzmann-Langevin dynamics of harmonic modes in nuclear matter
is analyzed within linear-response theory,
both with an elementary treatment
and by utilizing the frequency-dependent response function.
It is shown how the source terms agitating the modes
can be obtained from the basic \BL\ correlation kernel
by a simple projection onto the associated dual basis states,
which are proportional to the RPA amplitudes
and can be expressed explicitly.
The source terms for the correlated agitation of any two such modes
can then be extracted directly,
without consideration of the other modes.
This facilitates the analysis of collective modes in unstable matter
and makes it possible to asses the accuracy
of an approximate projection technique employed previously.

\vfill
{\small \noindent
$^*$This work was supported by the Director,
Office of Energy Research,
Office of High Energy and Nuclear Physics,
Nuclear Physics Division of the U.S. Department of Energy
under Contracts No.\ DE-AC03-76SF00098 and DE-FG05-89ER40530,
the Commission of the European Community under Contract No.\ ERBCHBI-CT-930619,
and the National Institute for Nuclear Theory
at the University of Washington in Seattle.
}

\end{titlepage}

\section{Introduction}
\label{intro}

Microscopic transport models are necessary
for the interpretation of heavy-ion collision experiments.
Approaches based on the nuclear Boltzmann equation
have been especially successful in accounting for a variety of observables,
including the collective flow pattern and the production of mesons
\cite{Schuck,Cassing}.
Models of this type seek to describe the one-body phase-space density
$f(\r,\p,t)$ of the nucleons in the system
(and of any other hadron species present).
The single-particle motion is governed by an effective Hamiltonian,
$h[f]=p^2/2m+U(\r)$,
and the corresponding collisionless evolution of $f$
is governed by the Vlasov equation \cite{Vlasov},
the semiclassical analog of the Time-Dependent Hartree-Fock equation
\cite{TDHF}.

The advantage of the semiclassical description is that
the constituent particles can be localized in phase-space
and so it is possible to take account of their residual short-range interaction
by means of a collision term,
as was first done by Boltzmann for dilute classical gases.
Due to the fermion nature of the nucleons,
the two-body scattering processes are suppressed by the blocking factors
$\bar{f}=1-f$ expressing the availability of a state.
This approach was first taken by Nordheim
for the study of electrons in solids \cite{N,UU}
and was later adapted for nuclear collision scenarios \cite{NPA314,BUU}.
This type of description is most often referred to as
the Boltzmann-Uhling-Uhlenbeck (\BUU) model \cite{guide}.

In the \BUU\ approach,
only the average effect of the collisions are taken into account.
While this suffices for many phenomena,
it is inadequate for processes exhibiting bifurcations,
such as nuclear fragmentation processes.
The description has therefore been extended to include
the fluctuating effect of the two-body collisions,
leading to the Boltzmann-Langevin (\BL) model \cite{Ayik}.
The equation of motion for $f$ is then of the form
\beq\label{BL}
{\del f\over\del t}
-\bfv\cdot{\del f\over\del\r}
+{\del U\over\del\r}\cdot{\del f\over\del\p}=
\bar{I}[f]+\delta I[f]
\eeq
where $\bar{I}(\r,\p,t)$ represents the average effect of the collisions
(the \BUU\ term),
while $\delta I(\r,\p,t)$ denotes the fluctuating remainder
(the Langevin term).

The presence of the stochastic Langevin term $\delta I$
in the equation of motion for $f$
causes continual branchings of the dynamical trajectories,
thus enabling the system to explore the entire range of possible fates and,
in particular, to exploit any instabilites encountered.
The evolution of the corresponding distribution of histories, $\phi[f](t)$,
can be described by a Fokker-Planck transport equation \cite{RR,BCR}.
The associated transport coefficients,
the drift coefficient $V$ and the diffusion coefficient $D$,
are given by
\beqar\label{V}
&~&V[f](\r,\p,t)\ =\ \prec \bar{I}(\r,\p,t) \succ\ ,\\
\label{D}
&~&2\hat{D}[f](\r,\p,\p',t)\delta(\r-\r')\delta(t-t')\ =\
\prec \delta I(\r,\p,t) \delta I(\r',\p,t')^* \succ\ ,
\eeqar
where $\prec\cdot\succ$ denotes an average over an ensemble of systems
that have all been prepared to have the same
one-body density $f$ at the time $t$.
The transport coefficients are useful
because they express how a given phase-space density $f(\r,\p)$
changes over a short period of time as a result of the collsions:
the drift coefficient is the average rate of change, $\prec\dot{f}\succ$,
while the diffusion coefficient expresses the early growth rate
of the correlation between the changes at the specified phase-space points,
$\dot{\sigma}(\r,\p;\r',\p')$.
Simple analytical approximations were recently derived
for these key quantities \cite{RA93}.

The Langevin extension of the nuclear Boltzmann model
is a significant advance,
since it makes it possible to address processes
in which fluctuations play a major role,
such as nuclear multifragmentation caused by the
irreversible development of unstable bulk modes \cite{PRL}.
To gain insight into this key process,
a recent study addressed the early evolution of nuclear matter
in the spinodal zone of the phase diagram \cite{CCR}.
The system is then mechanically unstable and the density fluctuations
generated by the Langevin term $\delta I$
may be amplified by the self-consistent effective field,
leading towards catastrophic transformations of the system
into an assembly of nuclear clusters.
That work developed a convenient formal framework
which provides instructive insight into the unstable dynamics
and makes it possible to obtain quantitative results.

It was shown that the collective modes are governed by simple feed-back
equations of motion,
in which the fluctuations generated by the Langevin term
are either magnified or suppressed by the adjusting effective field.
The agitation rate is given in term of source terms ${\cal D}$
for which approximate expressions were derived,
and they have been used to obtain quantitative results \cite{CCR,CC}.
It is especially important to know accurately the agitation rates
for the most unstable modes,
since these will tend to become dominant,
and in fact the final outcome may depend sensitively on their value.
For example,
an expanding system passing through the unstable spinodal zone
may multifragment if sufficiently large density fluctuations are
produced during the finite time available,
but if the fluctuations remain too small
the system may instead emerge relatively intact and then merely vaporize
\cite{CCGJ}.
Moreover,
if the fluctuation amplitude is well understood,
it may be possible to develop simplified stochastic mean-field approaches,
where the complicated Langevin term is replaced by a simpler approximate term
leading to the same agitation rates \cite{noise,BOB}.

Therefore,
in the present paper, we revisit this problem.
We shall first address the situation by elementary means,
thereby gaining instructive insight into the key problems,
in particular the nature of the approximate treatment
employed earlier \cite{CCR}.
Subsequently, we employ the frequency-dependent response function
to obtain the same result in a more direct manner.
As a main result,
we find the explicit form of the dual basis states
which makes it possible to obtain the collective source terms directly.
Furthermore,
the accuracy of the inital approximate treatment given in ref.\ \cite{CCR}
is illustrated.

\section{Elementary treatment}
\label{elem}

We are considering the early evolution of nuclear matter
that has been prepared with a uniform density distribution $\rho_0$
and with a Fermi-Dirac momentum distribution
corresponding to a specified temperature $T$.
It is therefore convenient to consider the Fourier transform
of the phase-space density,
\beq\label{F}
f_{\k}(\p,t)=\int{d\r\over\sqrt{\Omega}}\
\rme^{-i\k\cdot\r}\ \delta f(\r,\p,t)\ ,
\eeq
where $\delta f=f-f_0$ is the deviation of $f(t)$ from the initial density
$f_0$
which depends only on the energy $\eps$.
The system is confined within a volume $\Omega$,
with periodic boundary conditions imposed,
so that the wave number $\k$ takes on discrete values.
As long as we remain within the regime of linear response,
the dynamical problem decouples with respect to wave number,
and so we may limit our considerations to a single value of $\k$.

The equation of motion for these Fourier coefficients 
follows readily from (\ref{BL}),
\beq\label{EoM}
{\del\over\del t}f_{\k} +i\k\cdot\bfv f_{\k}
-i{\del U_k\over\del\rho} \k\cdot\bfv {\del f_0\over\del\eps}\rho_{\k}
=\delta I_\k\ ,
\eeq
Here $\del U_k/\del\rho$ represents the appropriate Fourier component
of the derivative of the effective field with respect to the density
and $\delta I_\k(\p,t)$ is the Fourier component
of the Langevin term in (\ref{BL}),
defined in analogy with eq.\ (\ref{F}).
Furthermore,
$\rho_\k(t)=\sum_\p f_\k(\p,t)$ is the Fourier transform of the matter density
$\rho(\r)=\sum_\p f(\r,\p)$,
where $\sum_\p$ denotes the summation over all the momentum states
and is equivalent to the integral $h^{-D}\int d\p$ in the limit of large
systems
having a continuum of momentum states.\footnote{We prefer to employ the
summation sign,
since the spectral structure is easier to visualize
when the problem is discretized by means of a lattice in momentum space.}
Finally,
we have ignored the average collision term $\bar I$,
since its effect is relatively small,
consisting primarily of a small reduction in the growth times
for the unstable modes \cite{Pethick,JR}.

The Vlasov equation (the left-hand part of (\ref{EoM})) has a complete set
of eigenfunctions,
\beq\label{fk}
f_\k^\nu(\p,t)=f_\k^\nu(\p)\ \rme^{-i\omega_\nu t}\ ,
\eeq
on which any time-dependent solution to (\ref{EoM}) may be expanded.
These eigenfunctions satisfy the equation
\beq
(\k\cdot\bfv-\omega_\nu) f_\k^\nu(\p)
={\del U_k\over\del\rho} \k\cdot\bfv {\del f_0\over\del\eps}\ \rho_\k^\nu\ ,
\eeq
and so they have the form
\beq
f_\k^\nu(\p)={\del U_k\over\del\rho}
{\k\cdot\bfv \over \k\cdot\bfv-\omega_\nu}{\del f_0\over\del\eps}\ ,
\eeq
where a specific normalization has been chosen,
$\rho_\k^\nu\equiv\sum_\p f_\k^\nu(\p)=1$.

The eigenfrequencies $\omega_\nu$
are determined by the dispersion relation,
\beq\label{disp}
1={\del U_k\over\del\rho}
\sum_\p {\k\cdot\bfv \over \k\cdot\bfv-\omega_\nu}{\del f_0\over\del\eps}
=-F_0 \chi(\omega_\nu)\ .
\eeq
In the last expression,
we have employed the usual Landau parameter $F_0$
and the free response function
\beq\label{chi}
\chi(\omega)\ =\
\sum_\p{\k\cdot\bfv\over\k\cdot\bfv-\omega}
{\del f_0\over\del\eps}/\sum_\p{\del f_0\over\del\eps}\ ,
\eeq
which is the average of $\k\cdot\bfv/(\k\cdot\bfv-\omega)$
over momentum space,
calculated with the weight function $\del f_0/\del\eps=-f_0\bar{f}_0/T$
emphasizing the Fermi surface.
It is easy to see that the solutions come in pairs of opposite signs
(which we may assign opposite values of the index $\nu$:
$\omega_{-\nu}=-\omega_\nu$).

In the absence of the damping term $\bar{I}$,
the dispersion relation (\ref{disp}) has exclusively real roots
as long as the system is prepared outside the spinodal zone,
\ie\ when $F_0>-1$.
However, inside the spinodal zone one pair of eigenvalues is purely imaginary,
$\omega_k=\pm i/t_k$, and is associated with a pair of collective modes,
one exponentially growing and the other one exponentially decaying,
with the same characteristic time constant $t_k$.
The amplified collective mode will quickly become the dominant one
and consequently this is the mode of primary concern.
However,
our present developments are quite general and apply to all the modes,
and they are not limited to the spinodal zone.

It is important to recognize that although the eigenfunctions $\{f_\k^\nu\}$
form a complete set, they are not orthogonal.
We therefore introduce the matrix $o_k^{\nu\nu'}$
as the inverse of the overlap matrix \cite{CCR},
\beq\label{o}
(o_k^{-1})^{\nu\nu'}=<f_\k^\nu|f_\k^{\nu'}>
\equiv \sum_\p f_\k^\nu(\p)^* f_\k^{\nu'}(\p)\ .
\eeq
Becuase of the rotational invariance,
the overlap matrix elements do not depend on the direction of
the wave vector $\k$ but only on its magnitude $k$.
Any solution to the equation of motion (\ref{EoM}), $f_\k(\p,t)$,
has a unique expansion on the eigenfunctions,
\beq\label{f}
f_\k(\p,t)=\sum_\nu A_\k^\nu(t) f_\k^\nu(\p)\ ,
\eeq
where the expansion coefficients are given by
\beq
A_\k^\nu=\sum_{\nu'} o_k^{\nu\nu'} <f_\k^{\nu'}|f_\k>\
=\ <q_\k^\nu|f_\k>\ .
\eeq
We have here introduced the functions
$q_\k^\nu(\p)\equiv\sum_{\nu'} o_k^{\nu\nu'} f_\k^{\nu'}(\p)$
which form the dual basis,
relative to the eigenfunctions $\{f_\k^\nu(\p)\}$, since
\beq
<q_\k^\nu|f_\k^{\nu'}>\ =\ \delta_{\nu\nu'}\ ,
\eeq
as is readily shown by using the definition (\ref{o}) of the overlap matrix.

Inserting the expansion (\ref{f}) into the equation of motion (\ref{EoM})
and projecting onto the dual basis 
we obtain the following equation for the expansion amplitudes,
\beq
{d\over dt}A_\k^\nu = -i\omega_\nu A_\k^\nu(t) +<q_\k^\nu|\delta I_\k(t)>\ .
\eeq
This form is easy to understand:
the noise term $\delta I$ acts as a source term that continually
produces stochastic changes of the amplitude $A_\k^\nu$,
while the Vlasov equation propagates the amplitude
in the associated effective field.
The above equation of motion can readily be solved formally,
\beq
A_\k^\nu(t)=\int_0^t dt'\
\rme^{i\omega_\nu(t'-t)}\ <q_\k^\nu|\delta I_\k(t')>\ +\ A_\k^\nu(0)\ .
\eeq

If we start from uniform matter,
the initial amplitudes vanish, $A_\k^\nu(0)=0$,
and since the noise term vanishes on the average,
$\prec \delta I \succ=0$,
the ensemble averages of the amplitudes remain zero,
$\prec A_\k^\nu(t)\succ=0$.
However, their correlations have a non-trivial evolution,
\beq\label{AA}
\prec A_\k^\nu(t) A_{\k'}^{\nu'}(t)^*\succ\ =\ \delta_{\k\k'}\
2{\cal D}_k^{\nu\nu'}
\int_0^t dt'\ \rme^{i(\omega_\nu-\omega_{\nu'}^*)(t'-t)}\ .
\eeq
We have here used the fact that the Langevin term is local in space and time,
so the problem decouples with respect to the wave number $\k$,
\beq
\prec \delta I_\k(\p,t) \delta I_{\k'}(\p',t')^*\succ\ =\
2\hat{D}(\p,\p')\ \delta(t-t')\ \delta_{\k\k'}\ .
\eeq
Furthermore,
we have introduced the source term
\beq\label{calD}
{\cal D}_k^{\nu\nu'}\ =\
<q_\k^\nu|D|q_\k^{\nu'}>\ \equiv\
\sum_{\p\p'} q_k(\p)^*\ \hat{D}(\p,\p')\ q_\k(\p')\ .
\eeq

We note that the above result (\ref{AA}) is equivalent to the following
equation of motion for the correlation coefficients
$\sigma_k^{\nu\nu'}(t)\equiv\prec A_\k^\nu(t) A_\k^{\nu'}(t)^*\succ$,
\beq
{d\over dt}{\sigma}_k^{\nu\nu'}(t)=2{\cal D}_k^{\nu\nu'}
-i(\omega_\nu-\omega_{\nu'}^*) \sigma_k^{\nu\nu'}(t)\ ,
\eeq
which is the result derived in ref.\ \cite{CCR}.
In order to better make contact with that work,
we note that the expression (\ref{calD}) for the source term
can be expressed in terms of the basis formed by the eigenfunctions,
\beq\label{Dexact}
{\cal D}_k^{\nu\nu'}\ =\
<{q}_\k^\nu|D|{q}_\k^{\nu'}>\ =\
\sum_{\mu\mu'} {o}_k^{\nu\mu}
<f_\k^\mu|D|f_\k^{\mu'}> {o}_k^{\mu'\nu'}\ .
\eeq
In ref.\ \cite{CCR} the focus was on the evolution of the two collective
modes in the spinodal zone and only those modes were retained.
So the full matrix $o_k^{\nu\nu'}$ was effectively replaced by
the $2\times2$ matrix $\tilde{o}_k^{\nu\nu'}$
involving only the two collective modes.
This amounts to approximating the exact dual state $q_\k^\nu(\p)$
by $\tilde{q}_\k^\nu(\p)\equiv\sum_{\nu'}\tilde{o}_k^{\nu\nu'}
f_\k^{\nu'}(\p)$,
where the sum extends only over the two collective modes.
The corresponding approximate expression for the collective source term is then
\beq\label{Dapprox}
\tilde{\cal D}_k^{\nu\nu'}\ =\
<\tilde{q}_\k^\nu|D|\tilde{q}_\k^{\nu'}>\ =\
\sum_{\mu\mu'} \tilde{o}_k^{\nu\mu}
<f_\k^\mu|D|f_\k^{\mu'}> \tilde{o}_k^{\mu'\nu'}\ ,
\eeq
as given in ref.\ \cite{CCR}.
This analysis exhibits the relationship between the two projection methods,
the initial one (\ref{Dapprox}) using the approximate dual basis functions
and the improved one (\ref{Dexact}) employing the exact dual basis.
In section \ref{disc} we illustrate this further by numerical comparisons.

\section{Response function approach}
\label{resp}

It is possible to treat the problem in an alternative manner,
by direct application of the response-function technique
\cite{Pethick,Lifshitz},
as is briefly summarized below.

Considering a general solution to (\ref{EoM}), $f_\k(\p,t)$,
we first perform a one-sided Fourier transform with respect to time,
\beq
f_\k(\p,\omega)=\int_0^\infty dt\ \rme^{i\omega t}\ f_\k(\p,t)\ ,
\eeq
leading to the equation
\beq\label{f_omega}
i(\k\cdot\bfv-\omega) f_\k(\p,\omega)
-i{\del U_k\over\del\rho} \k\cdot\bfv {\del f_0\over\del\eps}\rho_\k(\omega)
=\delta I_\k(\p,\omega)\ ,
\eeq
where $\delta I_\k(\p,\omega)$ is the corresponding transform of
$\delta I_\k(\p,t)$.

It is common to introduce the susceptibility
\beq
\vareps({\omega}) \equiv 1+F_0\chi(\omega)
=1-{\del U_k\over\del\rho}
\sum_\p {\k\cdot\bfv \over \k\cdot\bfv-\omega}{\del f_0\over\del\eps}\ ,
\eeq
which is simply related to the response function $\chi(\omega)$
introduced in (\ref{chi}).
The solution of the above equation (\ref{f_omega})
can then be written as a simple projection of the noise,
\beq
\rho_\k(\omega)\equiv\sum_\p f_\k(\p,\omega)=
-{i\over\vareps({\omega})}
\sum_\p{\delta I_\k(\p,\omega)\over \k\cdot\bfv-\omega}\ .
\eeq
The Fourier component of the density can subsequently be obtained
by applying the inverse transform,
\beq\label{inverse}
\rho_\k(t)=\int_C {d\omega\over2\pi}\ \rme^{-i\omega t}\ \rho_\k(\omega)
=\int_0^\infty dt' \int_C {d\omega\over2\pi i}\
{\rme^{i\omega(t'-t)}\over\vareps(\omega)}
\sum_\p {\delta I_\k(\p,t') \over \k\cdot\bfv-\omega}\ .
\eeq
Here the integral is to be carried out along a path $C$ in the complex plane
that passes above all the poles of the integrand,
from the far left to the far right.
When $t'>t$ the $\omega$ integral can be carried out by completing the contour
around the upper halfplane where the integrand vanishes rapidly.
Since the completed contour encloses no poles,
the $\omega$ integral vanishes
and so the $t'$ integration effectively extends only up to $t'=t$.
To evaluate the $\omega$ integral for $t'<t$,
we complete the contour around the lower halfplane,
which adds no contribution,
and then apply the Residue Theorem.
The poles are determined by the condition
$\vareps(\omega_\nu)=0$,
which is recognized as the dispersion relation (\ref{disp})
and so they occur at the eigenfrequencies, as expected.
Furthermore,
the residue of the susceptibiltiy is
\beq
\vareps'(\omega_\nu)\equiv
\left({\del\vareps\over\del\omega}\right)_{\omega=\omega_\nu}
=-\sum_\p{f_\k^\nu(\p)\over\k\cdot\bfv-\omega_\nu}
=-<Q_\k^\nu | f_\k^\nu>\ ,
\eeq
where we have introduced the auxiliary function
$Q_\k^\nu(\p)\equiv1/(\k\cdot\bfv-\omega_\nu^*)$,
which is recognized as the usual RPA amplitude.
We may then write the result (\ref{inverse}) on a compact form,
\beq\label{rhok}
\rho_\k(t)=\int_0^t dt'\sum_\nu \rme^{i\omega_\nu(t'-t)}\
{<Q_\k^\nu | \delta I_\k(t')> \over <Q_\k^\nu | f_\k^\nu>}
=\sum_\nu A_\k^\nu(t)\ .
\eeq
The last relation arises by recalling the expansion (\ref{f})
of the solution on the eigenfunctions,
and the fact that we use a normalization such that $\rho_\k^\nu$ is unity.

Because of the stochastic character of the evolution,
we need to consider the entire ensemble of possible dynamical histories.
The ensemble average of the Langevin noise term vanishes,
$\prec \delta I_\k(\p,t)\succ=0$,
and so $\prec A_\k^\nu(t)\succ=0$,
as in sect.\ \ref{elem}.
The corresponding correlation function is given by
\beq\label{sigmak}
\sigma_k^{\nu\nu'}(t)
={ <Q_\k^\nu|2D|Q_{\k}^{\nu'}> \over
 <Q_\k^\nu|f_\k^\nu> <f_{\k'}^{\nu'}|Q_{\k}^{\nu'}> }\
\int_0^t dt'\ \rme^{i(\omega_\nu-\omega_{\nu'}^*)(t'-t)}\ .
\eeq
The results (\ref{rhok}) and (\ref{sigmak}) are identical to
what was derived in sect.\ \ref{elem},
since the auxiliary function $Q_\k^\nu(\p)$
is proportional to the normalized dual basis function $q_\k^\nu(\p)$,
\beq\label{qQ}
q_\k^\nu(\p)={Q_\k^\nu(\p) \over <Q_\k^\nu|f_\k^\nu> }\ .
\eeq
This key relation can be verified
by using the dispersion relation (\ref{disp})
and the fact that the eigenvalues come in opposite pairs,
as is demonstrated in the Appendix.

The above derivation shows that
the Boltzmann-Langevin dynamics in nuclear matter
can be treated by suitable adaptation of
the standard response-function approach \cite{Pethick,Lifshitz}.
The availability of an explicit form of the dual basis
enables us to determine the source term ${\cal D}_k^{\nu\nu'}$ directly
for each pair of modes $\nu$ and $\nu'$,
without the need for invoking the entire overlap matrix $o_k^{\nu\nu'}$.
This is particularly useful inside the spinodal zone,
since the dynamics quickly becomes dominated by the amplified collective mode,
so the there is little need for considering the others.
Accordingly,
the dual projection method was employed in a recent study of
the effect of memory time on the agitation of unstable modes in nuclear mattter
\cite{AR94}.

\section{Discussion}
\label{disc}

We have shown how the Boltzmann-Langevin dynamics of nuclear matter
can be treated within linear response theory.
This treatment is valid as long as the induced deviations remain small
and so, for matter in the unstable phase region,
it can be used to understand the onset of the spinodal decomposition.

The time evolution of the density undulations can be separated
into two qualitatively different stages.
The characteristic time separating the two dynamical regimes is given by the
amplification time $t_k$.
Whereas all the modes are initially agitated in a rather democratic manner,
in accordance with the miscoscopic diffusion coefficient $\hat{D}(\p_1,\p_2)$,
only the amplified collective mode will ultimately dominate
(since the suppressed collective mode will saturate
and the non-collective ones have no exponential development).
Therefore,
if one is interested in the long-term behavior,
\ie\ the appearance of the system after a time longer than
the characteristic time,
then it suffices to retain only the source term
for the amplified collective mode.
That is why we have focussed on the dynamics of the collective modes.
The agitation rate for the amplified collective mode
is most easily obtained by projecting the diffusion coefficient
$\hat{D}(\p_1,\p_2)$ onto the dual basis state for that mode, $q_\k^+(\p)$,
as indicated in eq.\ (\ref{sigmak}).

However,
it should be recognized that for times short in comparison with the
amplification time, $t<t_k$,
the relative magnitude of the contributions
from other modes is not yet negligible.
This is most clearly brought out by the fact that
the non-collective contributions cancel exactly the collective ones
to leading order in time,
so that $\sigma_k\sim t^2$ at first,
whereas the expression (\ref{sigmak}) yields a linear initial growth.
This general property follows formally from particle-number conservation
which dictates that the integral of the microscopic diffusion coefficient
with respect to momentum vanish.
The feature can easily be understood from the fact that the
collision term is local in space and time.
It then produces merely a rearrangement of the local occupation
in momentum space, without affecting directly the matter density,
and fluctuations in the matter density only appear subsequently
as the generated two-particle-two-hole excitons are propagated
in the effective field.
Consequently the effect of the Langevin term on the matter density
is only second order in time.
Contrary to this,
the density fluctuation arising from the agitation of the amplified mode alone
is initially linear in time, $\sigma_k^{++}\approx2{\cal D}_k^{++}t$.

The key quantities, such as the collective source terms,
can be evaluated either by direct numerical calculation \cite{CCR,CC}
or by employing the analytical approximation
derived for the diffusion coeffision in ref.\ \cite{RA93}
and making use of the dispersion relation.
For our present illustrative purposes,
we follow the latter approach,
which is discussed in detail in ref.\ \cite{JR}
and employs a realistic two-body interaction \cite{Emil1}
for the calculation of the Landau parameter $F_0$.

Let us first recall that the spinodal zone
is situated within a parabola-like boundary
that extends from zero density to about two thirds of the saturation density
for vanishing temperature and reaches upwards to the critical temperature
of about 16 MeV for symmetric and uncharged nuclear matter.
This standard spinodal boundary pertains to undulations
with a vanishing wave number, $k=0$,
and it shrinks as the wave number is increased,
and finally disappear altogether at a point located on the  $T=0$ axis
near one third of the saturation density.
The amplification time $t_k(\rho,T)$ is infinitely large at the boundary
and becomes shorter as the phase point considered
is moved closer towards that limiting point.
The fastest amplification time is about $0.7\cdot10^{-22}\ \rm s$
but typical values are somewhat larger,
due to the finite temperature.

Figure \ref{f:1} shows the collective source term ${\cal D}_k^{++}$
obtained with the model presented in ref.\ \cite{JR}.
We note that this quantity diverges at the spinodal boundary where
$F_0\rightarrow-1$,
which may serve as a reminder of the limitations mentioned above.
Near the boundary the characteristic time tends to infinity,
$t_k\rightarrow\infty$,
and so we remain in the early dynamical regime where the other modes
are significant;
indeed, they will conspire to cancel exactly the divergent behavior
of the amplified mode.

The original orthogonal projection introduced in ref.\ \cite{CCR}
effectively employs an approximate dual basis function
obtained by decoupling the $2\times2$ collective part of the overlap matrix
from the rest.
The result is shown in fig.\ \ref{f:1} by the dashed curves
and it is seen to behave in a manner very similar to
that obtained by projecting onto the exact dual basis function (solid curves).
Indeed, it exhibits a similar divergent behavior near the spinodal boundary.
However, it is typically smaller by about 30\%
for the most rapidly amplified modes,
and so the use of the exact dual basis represents a
significant quantitative improvement.

In order to compare the two methods with regard to their predictions
for the density fluctuations,
we consider the Fourier component of the density variance,
\beqar	\label{sigmat}	\nonumber
\sigma_k(t)&\equiv&
\int d\r_{12}\ \rme^{-i\k\cdot(\r_1-\r_2)}
\prec\delta\rho(\r_1)\delta\rho(\r_2)\succ\\
&=&\prec\rho_\k^*\rho_\k\succ\
\approx\ 2{\cal D}_k^{++} t_k \sinh({2t\over t_k}) + 4{\cal D}_k^{+-}t\ ,
\eeqar
where the last expression has been obtained by retaining only
the two collective modes,
and it has been used that ${\cal D}_k^{++}={\cal D}_k^{--}$
and ${\cal D}_k^{+-}={\cal D}_k^{-+}$.
The resulting variance is displayed in fig.\ \ref{f:2}
for a sequence of times $t$.
The approximate projection method involves the inversion of the
$2\times2$ collective overlap matrix.
As the spinodal boundary is approached,
the collective frequency tends to zero
and the two collective modes become identical.
The mixed source term then also diverges and, as it happens,
$\tilde{\cal D}_k^{++}+\tilde{\cal D}_k^{+-}\sim t_k^{-1}$
as $t_k\rightarrow\infty$.
So the resulting density variance $\sigma_k$ tends to zero.
Such a regular behavior does not arise when the
exact projection is employed,
because the mixed source term remains regular and so cannot cancel the
divergence of the diagonal term.
The resulting density variance then diverges at the spinodal boundary,
emphasizing the fact that the method is limited to scenarios
well within the spinodal zone where the characteristic time is reasonably
short.

In fig.\ \ref{f:3} we show the density variance $\sigma_k$
as a function of the wave number $k=2\pi/\lambda$
and at the same times,
for the typical temperature $T=4\ \MeV$
and for a density near which the fastest amplification occurs,
$\rho=0.3\rho_0$.
It is seen how those modes that have the shortest amplification time
grow progressively dominant,
so that the density fluctuations emerge with a rather narrow Fourier spectrum
centered around the fastest-growing mode associated with the specified
density and temperature.
It should be noted that the exponential growth limits the length of time
over which the linear-response treatment remains valid.
{}From the results shown in fig.\ \ref{f:3} it can readily be estimated
that the predicted average magnitude of the density fluctuation
equals the average density for $t\approx3.5\cdot10^{-22}\ {\rm s}$,
so the results are not expected to be meaningful beyond
$t\approx3\cdot10^{-22}\ {\rm s}$.
A comparison between the results of the two different projection methods
leads to the same conclusion as reached above:
use of the approximate dual basis states
reduces the density variances by about 30\%,
except near the spinodal boundary where the characteristic time diverges
and it becomes insufficient to consider only the collective modes.

While we expect that the projection method employing the exact dual basis
states
is quite reliable for times exceeding the respective amplification time,
$t>t_k$,
it still needs to be determined how accurate it is at earlier times.
In particular,
it would be of interest to improve the projection method
so that it is applicable also near the spinodal boundary.
This task is of practical importance,
since the systems prepared in heavy-ion collisions
are initially situated outside the spinodal region
and so must cross the boundary to become unstable.

{\bf Acknowledgements:}
This work was supported by the Director,
Office of Energy Research,
Office of High Energy and Nuclear Physics,
Nuclear Physics Division of the U.S. Department of Energy
under Contracts No.\ DE-AC03-76SF00098 and DE-FG05-89ER40530 and by
the Commission of the European Community under Contract No.\ ERBCHBI-CT-930619.
The authors also wish to acknowledge support and hospitality
by the Nuclear Theory Group at the Lawrence Berkeley Laboratory
(SA, PC, and MC),
the National Institute for Nuclear Theory
at the University of Washington in Seattle (SA and JR),
and GANIL (SA)
while this work was carried out.

\appendix
\section{The explicit form of the dual basis}

In this Appendix we prove that the auxiliary functions
$Q_\k^\nu(\p)\sim1/(\k\cdot\bfv-\omega_\nu^*)$
are indeed proportional to the dual basis $q_\k^\nu(\p)$
characterized by $<q_\k^\nu|f_\k^{\nu'}>=\delta_{\nu\nu'}$.
We first recall that the eigenvalues $\omega_\nu$ come in pairs of opposite
sign
and that we label them accordingly, $\omega_{-\nu}=-\omega_\nu$.
Let us start with the most frequent case
when the two eigenvalues differ, $\omega_\nu\neq\omega_{\nu'}$,
\beqar
&~&<Q_\k^\nu|f_\k^{\nu'}> = \sum_\p {1\over \k\cdot\bfv-\omega_\nu}
{\k\cdot\bfv\over \k\cdot\bfv-\omega_{\nu'}} {\del f_0\over\del\eps}\\
&~&={1\over \omega_\nu-\omega_{\nu'}}
\left( \sum_\p{\k\cdot\bfv \over \k\cdot\bfv-\omega_\nu}
{\del f_0\over\del\eps}
-\sum_\p{\k\cdot\bfv \over \k\cdot\bfv-\omega_{\nu'}}
{\del f_0\over\del\eps} \right)=0\ ,
\eeqar
This result is obtained after a few of elementary manipulations,
followed by application of the dispersion relation (\ref{disp})
to each of the two terms, which then cancel.
This leaves only the possiblity the the two eigenvalues are equal,
$\nu=\nu'$,
in which case we have
\beq
<Q_\k^\nu|f_\k^{\nu'}> = \sum_\p {1\over \k\cdot\bfv-\omega_\nu}
{\k\cdot\bfv\over \k\cdot\bfv-\omega_\nu} {\del f_0\over\del\eps}
=\sum_\p {2\omega_\nu (\k\cdot\bfv)^2 \over ((\k\cdot\bfv)^2-\omega_\nu^2)^2}
{\del f_0\over\del\eps} \neq0\ .
\eeq
Therefore, after a suitable renormalization, $q_\k^\nu={\cal N}_\k Q_\k^\nu$,
the required orthonormality relation follows,
$<q_\k^\nu|f_\k^{\nu'}>=\delta_{\nu\nu'}$,
\ie\ $\{q_\k^\nu\}$ indeed forms the dual basis.

\bfig
\caption{Comparison of source terms.}
\label{f:1}
The source term ${\cal D}_k^{++}$ for the amplified collective mode
having a wave length $\lambda=2\pi/k$,
in nuclear matter prepared with a uniform density $\rho$
and with a specified temperature $T$.
The solid curves are obtained by projecting onto the exact dual basis state,
eq.\ (\ref{Dexact}),
while the dashed curves show the corresponding results
obtained with the approximate dual basis state, eq.\ (\ref{Dapprox}).
The calculations have been done with the approximate formulas
developed in ref.\ \cite{JR}.
The upper panel considers the typical temperature $T=4\ \MeV$
and illustrates the dependence on the wave length $\lambda$,
while the lower panel keeps the wave length fixed at $\lambda=2\pi/k=8\ \fm$,
near which the most rapid amplification occurs,
and illustrates the temperature dependence.
\efig

\bfig
\caption{Density fluctuations.}
\label{f:2}
The variance of the density fluctuations associated with the collective modes,
$\sigma_k$,
after a given time $t=1,2,3,4\cdot10^{-22}\ \rm s$ has elapsed,
as a function of the average density
and for the temperature $T=4\ \MeV$ and a wave length of $\lambda=8\ \fm$.
Notation and model are as in fig.\ \ref{f:1}.
\efig

\bfig
\caption{Dependence on wave number.}
\label{f:3}
The variance of the density fluctuations associated with the collective modes,
$\sigma_k$,
after a given time $t=1,2,3,4\cdot10^{-22}\ \rm s$ has elapsed,
as a function of the wave number $k=2\pi/\lambda$,
for the typical temperature $T=4\ \MeV$
and for the density $\rho=0.3\rho_0$.
Otherwise similar to fig.\ \ref{f:2}.
\efig


\newpage
\begin{thebibliography}{30}

\bibitem{Schuck}
P. Schuck, R.W. Hasse, J. J\"anicke, C. Gregoire, B. Remaud, F. Sebille,
and E. Suraud, Prog. Part. Nucl. Phys. 22 (1989) 181

\bibitem{Cassing}
W. Cassing, V. Metag, U. Mosel, and K. Niita,
Phys. Rep. 188 (1990) 363

\bibitem{Vlasov}
W. Cassing and U. Mosel, Prog. Part. Nucl. Phys. 25 (1990) 235

\bibitem{TDHF}
P. Bonche, S.E. Koonin, and J.W. Negele, Phys. Rev. C13 (1979) 226

\bibitem{N}
L.W. Nordheim, Proc. Roy. Soc. A119 (1928) 689

\bibitem{UU}
E.A. \"Uhling and G.E. Uhlenbeck, Phys. Rev 43 (1933) 552

\bibitem{NPA314}
J. Randrup, Nucl. Phys. A314 (1979) 429

\bibitem{BUU}
G.F. Bertsch, H. Kruse, and S. Das Gupta, Phys. Rev. C (1984) 673

\bibitem{guide}
G.F. Bertsch and S. Das Gupta, Phys. Rep. 160 (1988) 190

\bibitem{Ayik}
S. Ayik and C. Gregoire,
Phys. Lett. B212 (1988) 269; Nucl. Phys. A513 (1990) 187

\bibitem{RR}
J. Randrup and B. Remaud, Nucl. Phys. A514 (1990) 339

\bibitem{BCR}
Ph. Chomaz, G.F. Burgio, and J. Randrup, Phys. Lett. B254 (1991) 340;
G.F. Burgio, Ph. Chomaz, and J. Randrup, Nucl. Phys. A529 (1991) 157

\bibitem{RA93}
J. Randrup and S. Ayik, Nucl. Phys. 572 (1994) 489

\bibitem{PRL}
G.F. Burgio, Ph. Chomaz and J. Randrup, Phys. Rev. Lett. 69 (1992) 885

\bibitem{CCR}
M. Colonna, Ph. Chomaz, and J. Randrup, Nucl. Phys. A567 (1994) 637

\bibitem{CC}
M. Colonna and Ph. Chomaz, Phys. Rev. C49 (1994) 1908

\bibitem{CCGJ}
M. Colonna, Ph. Chomaz, A. Guarnera, and B. Jacquot, in preparation

\bibitem{noise}
M. Colonna \etal, Phys. Rev. C47 (1993) 1395

\bibitem{BOB}
Ph. Chomaz, M. Colonna, A. Guarnera, and J. Randrup,
LBL-35988 (1994), Phys. Rev. Lett. (in press)

\bibitem{Pethick}
C.J. Pethick and D.G. Ravenhall, Ann. Phys. 183 (1988) 131

\bibitem{JR}
J. Randrup, LBL-35848 (1994), in preparation

\bibitem{Emil1}
E. de Lima Medeiros and J. Randrup, Nucl. Phys. A529 (1991) 115

\bibitem{Lifshitz}
E.M. Lifshitz and L.P. Pitaevskii, {\em Physical Kinetics},
Pergamon Press, New Your (1981)

\bibitem{AR94}
S. Ayik and J. Randrup, LBL-35745 (1994), Phys. Rev. C (in press)

\end{thebibliography}
\end{document}